\documentclass[%
aps,
pre,
twocolumn,
floatfix,
superscriptaddress,
groupedaddress,
 amsmath,amssymb,
]{revtex4}

\usepackage{graphicx}% Include figure files
\usepackage{dcolumn}% Align table columns on decimal point
\usepackage{bm}% bold math
\usepackage[utf8]{inputenc}
\usepackage[artemisia]{textgreek}
\usepackage{float}

\begin{document}

%\preprint{APS/123-QED}

\title{Gene regulatory and signalling networks exhibit distinct topological distributions of motifs}% Force line breaks with \\

\author{Gustavo Rodrigues Ferreira}
\affiliation{São Carlos Institute of Physics, University of São Paulo, São Carlos, Brazil}

 \author{Helder Imoto Nakaya}
\affiliation{School of Pharmaceutical Sciences, University of São Paulo, São Paulo, Brazil}
 
\author{Luciano da Fontoura Costa}%
 \email{ldfcosta@gmail.com}
\affiliation{São Carlos Institute of Physics, University of São Paulo, São Carlos, Brazil}

\date{\today}% It is always \today, today,

\begin{abstract}
The biological processes of cellular decision making and differentiation involve a plethora of signalling pathways and gene regulatory circuits. These networks, in their turn, exhibit a multitude of motifs playing crucial parts in regulating network activity. Here, we compare the topological placement of motifs in gene regulatory and signalling networks and find that it suggests different evolutionary strategies in motif distribution for distinct cellular subnetworks.
%\begin{description}
%\end{description}
\end{abstract}

\pacs{Valid PACS appear here}% PACS, the Physics and Astronomy
\maketitle

%\tableofcontents

\section{\label{sec:int}Introduction}
In mathematics, cellular biological processes can be represented through concepts of graph theory \cite{Barabasi,Alon,Buchanan,Cell}. In these models, proteins and genes are depicted as nodes, and the chemical reactions or regulatory interactions between them as edges. Such representations of molecular systems, called networks, highlight the extensive crosstalk between a cell's components and the complex ways in which it regulates itself. To function, cells recruit or silence specific subsets of nodes and edges that have to be both spatially and temporally coordinated \cite{Boris}.

\begin{figure}[!h]
    \centering
    \includegraphics[width=.95\columnwidth]{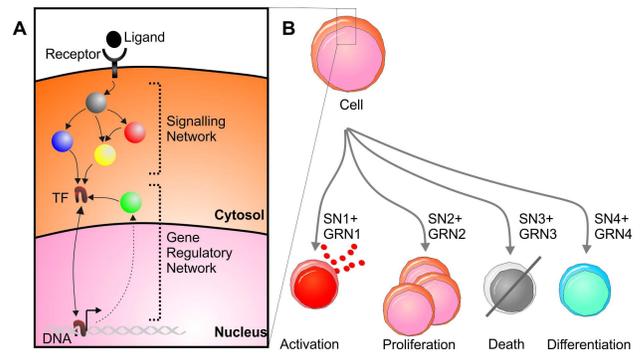}
    \caption{Signal transduction in eukaryotic cells. The signalling cascade initiated by a ligand-receptor binding (A) drives cellular decision-taking depending on the activated receptor (B).}
    \label{fig:sigpath}
\end{figure}

This distribution inside the cell serves the purpose of integrating and propagating hundreds of distinct signals and stimuli. This complex process, known as signal transduction, involves two different types of networks (Figure \ref{fig:sigpath}A): signalling networks (SNs) in the cytosol and gene regulatory networks (GRNs) in the nucleus. The former consists of a series of biochemical reactions that activate or inactivate proteins, channels and transcription factors -- generally starting with the binding of a ligand molecule to a receptor protein. In an SN, nodes are biochemical species that undergo the aforementioned reactions, and an edge from species X to Y indicates that X triggers or ends the activity of Y.

The GRN is a network composed of transcription factors (proteins) that enhance or inhibit the translation of other genes, including themselves. Signal propagation in a GRN usually initiates with the translocation of an activated transcription factor to the nucleus, where it activates the transcription of specific targets (see Figure \ref{fig:sigpath}A). Both kinds of networks show a common theme: signal propagation starts from a specific origin -- which we dub a receptor, input, or upstream node.

The different combinations of activated SNs and GRNs will determine the cell's response to one or more stimuli (see Figure \ref{fig:sigpath}B). In fact, different activation patterns for the same receptor can also induce distinct cellular responses \cite{Tomida,LiuX}, making for an extremely diverse signal processing system. For instance, ligands such as EGF and TGF-\textbeta{} induce cell proliferation \cite{Pardee}, whereas IFN-\textgamma{} and IL-4 induce B lymphocytes to secrete antibodies \cite{Nutt}.

Thus, understanding the networks' topology, or structure, is crucial to grasping the pathway's qualitative responses. For instance, a cell's ability to endure deleterious mutations in transcription factors has been shown to evolve gradually with changes in its GRN's topology \cite{RobTop}. Its resilience to total collapse (here meaning a sudden, irreversible transition to a state where all cellular activity ceases) is also known to be defined by a delicate interplay between its biochemical parameters and GRN structure \cite{resBar}.

\begin{figure}[!h]
    \centering
    \includegraphics[width=.5\columnwidth]{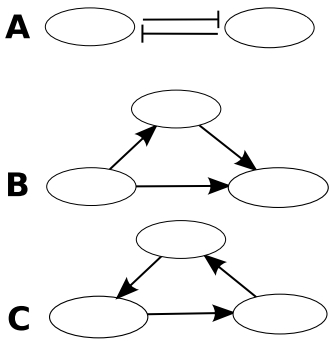}
    \caption{Examples of motifs found in cellular systems: the switch (A), the feed-forward loop, or FFL (A), and the feedback loop, or FBL (C). The names refer to specific instances of these structures observed in real cellular networks, and stand for transcription factors participating in cell development and differentiation.}
    \label{fig:mots}
\end{figure}

Of particular interest, then, are certain ubiquitous interaction patterns (or subgraphs), which have been termed motifs \cite{Alon,Milo}. The switch, the feed-forward motif and the feedback loop, shown in Figure \ref{fig:mots}, are examples of this class, and all of them have been shown to possess special dynamical properties regarding signal transduction and transmission \cite{SysBio,Ghaff,Shin,Liu}.

Previous works based on this premise have been successful in identifying signalling motifs that participate in signal transduction and cellular decision making \cite{Diella}. Furthermore, it was shown \cite{Maayan} that motifs tend to organise themselves in clear regions around cell receptors, suggesting both a role in signal processing and the importance of their precise topological placement in SNs. However, the topological distribution of motifs in GRNs has yet to be compared to this SN-minded paradigm.

In this paper, we compared the distribution of motifs in both the GRNs and SNs to assess their roles in signal propagation during cell differentiation. We used publicly available data from online databases to construct the networks and analysed them using local concepts from network theory, emphasising the characterisation of motifs along topological neighbourhoods \cite{Maayan,Filipi,Hier}. The results indicate two main types of organisation. The motifs in SNs tended to organise in symmetric, concentric layers around the receptor. On the other hand, in the case of GRNs, the motifs spread out in an asymmetric fashion along the hierarchical layers.

\section{\label{sec:methods}Methods}
\subsection{\label{sec:data}Network Construction}
The two types of networks analysed here, SNs and GRNs, have been made available in public databases. For the GRNs, the RegNetwork \cite{RegNetwork} database is a knowledge-based collection of regulatory interactions between transcription factors, microRNAs (miRNAs) and target genes. It combines and synthesises information from curated databases such as GenBank, BioGrid and Ensembl.

The signalling pathways were taken from Reactome \cite{Reactome} and processed with the rBiopaxParser package \cite{rbiopax}. This processing entails extracting the network structure from a Biopax annotation file, used in such public databases, and converting it to an edge list suited to our needs. Reactome is a database of curated interactions maintained by a collaboration among several research institutes, and also integrates orthology-based information from Ensembl.

For each signal transduction pathway considered here, a GRN subnetwork consisting of the genes involved in the associated biological process was extracted. This serves the double purpose of reducing the computational load of our analyses and focusing our attention on the biological entities that are actually relevant to the process.

More specifically, for each pathway, an RNA sequencing (RNA-seq) study concerning an associated differentiation process was used to identify Differentially Expressed Genes (DEGs). The pre-processed expression profiles were taken from the Sequence Read Archive \cite{SRA} (SRA), and DEGs were selected in a two-stage process: variance control and subsequent filtering of highly variant genes, as described below.

In an RNA-seq expression profiling, gene expression (under a null hypothesis of no change between samples) is modelled as a negative binomial distribution \cite{deseq}. This distribution is characterised by a strong dependence between mean and variance, which affects subsequent variance-based analyses. Thus, the expression profiling is transformed according to the Variance Stabilising Transformation for negative binomial data, derived by Guan \cite{VST}. More specifically, Guan proves that if $X$ is a negative-binomial-distributed random variable with parameters $r$ and $p$, then the transformed variable $Y = \sqrt{r} \sinh^{-1} \sqrt{X/r}$ has no dependence between mean and variance.

After removing mean-variance dependence, variance between different sample phenotypes (obtained from the SRA) was used as a criterion for DEG selection. We performed an Analysis of Variance (ANOVA; \cite{ANOVA}), which assesses the statistical significance of the within-class variance to total variance ratio, and attributed a p-value to each gene. Then, we controlled the rate of false positive discoveries with Benjamini and Hochberg's False Discovery Rate (FDR; \cite{FDR}) method. The FDR orders the p-values increasingly, so that $p_1 \leq p_2 \leq \ldots \leq p_n$, and selects those satisfying $p_k \leq \alpha k/m$, where $\alpha$ is the desired p-value threshold for an individual test. It is known (see, for instance, \cite{Giraud}) that this procedure has an expected False Discovery Rate upper bounded by $\alpha$.

\subsection{\label{sec:netana}Network Analysis}
A central concept to our analyses is that of a node's neighbourhood \cite{Hier}. Given a network, represented as a graph $G$, and a node $v$ in $G$, the $d$-th neighbourhood of $v$, denoted $R_d(v)$, is the set of all nodes accessible in at most $d$ steps from $v$. In a cellular network, successive neighbourhoods act as a proxy for the dynamics of signal propagation in the cell; since edges represent direct regulatory interactions, the chemical reactions associated to signal transduction occur along paths in the cellular network.

Associated to a node's neighbourhoods are several different measures. Of interest to us are the concepts of concentric symmetry and motif cumulative distribution (mCDF). The former is defined as the entropy associated to transition probabilities of a walk $h$ steps long starting from $v$ \cite{Filipi}. In other words, it quantifies how similar the possible walks of a certain length around a node can be -- the higher its value, the more similar they are, and thus the more symmetric the node's neighbourhood. It is preferred here over the automorphism-related symmetry metrics \cite{Biggs,Xiao} due to the latter's computational intractability (in general) and poor normalisation (automorphism groups have a loose upper bound on their order at around $N!$ for a graph on $N$ nodes, as discussed by Silva et al. \cite{Filipi}).

The motif cumulative distribution addresses how a certain motif is arranged in the neighbourhoods of a node. Given a motif structure, a node $v$ and a graph $G$, it is defined as the amount of occurrences of the motif along successive neighbourhoods $R_d(v)$, subsequently normalised as to approach unity.

The mCDF is tied to the Motif Location Index (MLI) metric used by Ma'ayan et al \cite{Maayan}. For a particular instance of a motif in a network, the MLI represents where the instance lies between a membrane receptor and a given cellular machinery. More specifically, both assess the placement of motifs relative to an origin (the receptor) in a cellular network. However, our work does not focus on specific cellular machines or processes, looking instead at generic signalling pathways. Thus, our metric lacks the ``distance to cellular machinery'' component of Ma'ayan's MLI, and thus reflects the positioning of motifs without regard to a particular cellular task (for our purposes, a cellular machinery is a subset of a GRN or SN containing nodes related to a particular cellular function, e.g. translation or cell division). This relation will be crucial when comparing our results to Ma'ayan's in section \ref{sec:res}.

Another important issue is determining whether the observed distributions are relevant when compared to randomly generated networks with the same degree distribution \cite{Sah,Stumpf}. We address this question by random sampling of networks through the edge-switching Monte Carlo algorithm described by Gkantsidis, Mihail and Zegura \cite{ESMC}. Different motif distributions were compared using the supremum distance for function spaces: if $f, g$ are two real-valued functions defined on $X$, the distance between them is $d(f, g) := \text{sup}_{x\in X} |f(x) - g(x)|$.

Based on this distance, the distributions $f_1, f_2, \ldots, f_n$ obtained through the sampling procedure, and given the observed distribution $f_{obs}$, we define a z-score $Z_{obs} = d(f_{obs}, \bar{f})/s_f$, where $\bar{f} := (1/n)\left(\sum_{i=1}^n f_i\right)$ and $s_f$ is the sample's standard deviation (in terms of the supremum distance). We also used a bootstrap p-value based on the distance to the mean; basically, $p$ is the proportion of times when $d(f_i, \bar{f}) > d(f_{obs}, \bar{f})$.

Another metric we incorporate from Ma'ayan et al. is the Density of Information Processing (DIP). It is defined as the increase in the amount of motifs divided by the increase in edges between consecutive neighbourhoods. In Ma'ayan et al., this ratio is multiplied by the grid coefficient \cite{Maayan,GC}, a generalisation of the clustering coefficient taking into account the formation of rectangles. However, since our considered motifs are of size smaller than four, we opted to forego this ``normalisation by grid coefficient''. In keeping with the idea that motifs are a network's processing units, this measure indicates the proportion of signal processing activity as information propagates through the network's paths.

\section{\label{sec:res}Results and Discussion}
We chose as representatives GRNs and SNs associated to three major signal transduction pathways in mouse (\textit{Mus musculus}) and human (\textit{Homo sapiens}). For the former, we studied the TCR signalling network, which drives the differentiation of T lymphocytes \cite{Becca}, and the EGF receptor (EGFR) pathway, involved in cell growth and survival \cite{Oda}. In humans, we studies the TGF-\textbeta{} pathway, which regulates cell growth, differentiation and apoptosis \cite{Shier,Massague}. The data for each pathway were taken from SRA as descried in section \ref{sec:data}. The RNA-seq profilings used were, respectively, GSE48138 \cite{Hu}, GSE86467 \cite{Fujii} and GSE36552 \cite{Yan}.

%According to our framework, each studied pathway presents two signal sources: a primary one, a membrane receptor in the SN component, and a secondary one, a transcription factor in the GRN component. For our chosen TCR, EGF and TGF-\textbeta{} pathways, the membrane receptors (as given by the Reactome database) are CSK, EGFR and TGFB1R, respectively. The associated transcription factors are, in the same order, Nfkb2, Junb and SMAD2. Since our focus is on the general structure of the networks, and not on the biomolecules themselves, 

The obtained networks were analysed with respect to their size (see Table \ref{tab:nets}) and degree distributions. We note, as expected, that signalling components are much smaller than their GRN counterparts. In keeping with the current literature \cite{Barabasi,RegNetwork}, degree distributions were seen to be power laws as determined by fitting a linear model to the log transformed degrees and degree probabilities -- all linear models had a coefficient of determination above $0.7$.

\begin{table}[!h]
\centering
\caption{Number of nodes and edges for our representative networks. Numbers on table are displayed as (nodes; edges).}
\label{tab:nets}
\begin{tabular}{|l|c|c|}
\hline
     & \textbf{SN} & \textbf{GRN} \\
     \hline
    \textbf{TCR} & 148; 848 & 3,835; 13,390\\
    \hline
    \textbf{EGFR} & 253; 1153 & 8,092; 36,266 \\
    \hline
    \textbf{TGF-\textbeta} & 67; 233 & 2,261; 5,384\\
\hline
\end{tabular}
\end{table}

%The three distinct pathways, each with two different networks (a cytosolic signalling pathway component and a GRN component), were analysed using both the motif cumulative distribution and the concentric symmetry metrics. For the motif cumulative distribution, our results show that signalling networks employ their motifs in a much more distinct fashion (``length-wise'') than GRNs (see Figure \ref{fig:MoCDF}). Additionally, the feedback loop (indicated by FBL) is given a special place in all networks, being placed further from the receptor node than other motifs. This tendency, already observed in signalling networks by Ma'ayan et al. \cite{Maayan}, is reinforced here and extended to gene regulatory circuitry. As with other principles of cell network evolution, this has plausibly come to be due to distinct dynamical properties for the feedback loop when compared to other motifs. We cite here the roles of feed-forward and feedback loops amplifying signals and filtering out noise, respectively, as an example \cite{SysBio,Shin}.

\begin{figure}[!h]
    \centering
    \includegraphics[width=.95\columnwidth]{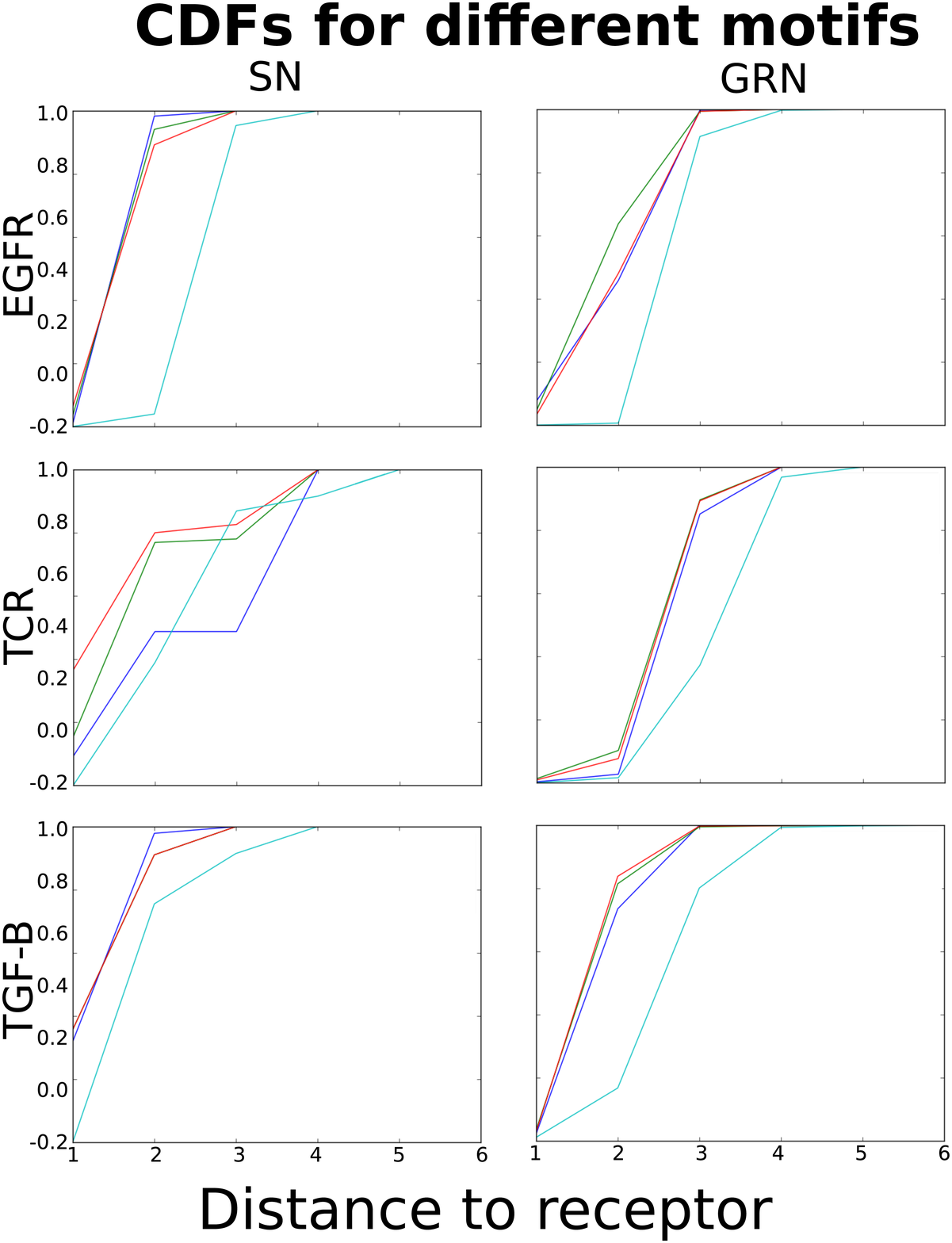}
    \caption{CDFs for distinct motifs in the various networks. Each colour represents a motif, as follows. Blue: feed-forward loop; green: double feed-forward loop; red: switch; light-blue (cyan): feedback loop.}
    \label{fig:MoCDF}
\end{figure}

The three distinct pathways, each with two different networks (a cytosolic signalling pathway component and a gene regulatory component), were analysed with respect to the mCDF, concentric symmetry and DIP metrics. For the motif cumulative distribution, our results show that signalling networks employ their motifs in a much more distinct fashion (``length-wise'') than GRNs (see Figure \ref{fig:MoCDF}). That is to say, different motifs may appear more strongly in different neighbourhoods (like in the TCR signalling network, Figure \ref{fig:MoCDF}, centre-left), or a certain motif may not be present in the network (notice how the double feed-forward loop is missing from the TGF-\textbeta{} signalling network, lower left). This variety of patterns in motif placement may have come about in two opposite ways: it might indicate that the placement of motifs in signalling networks is not under any evolutionary constraint, or it might suggest tailored distributions for each pathway.

\begin{figure}[!h]
    \centering
    \includegraphics[width=.95\columnwidth]{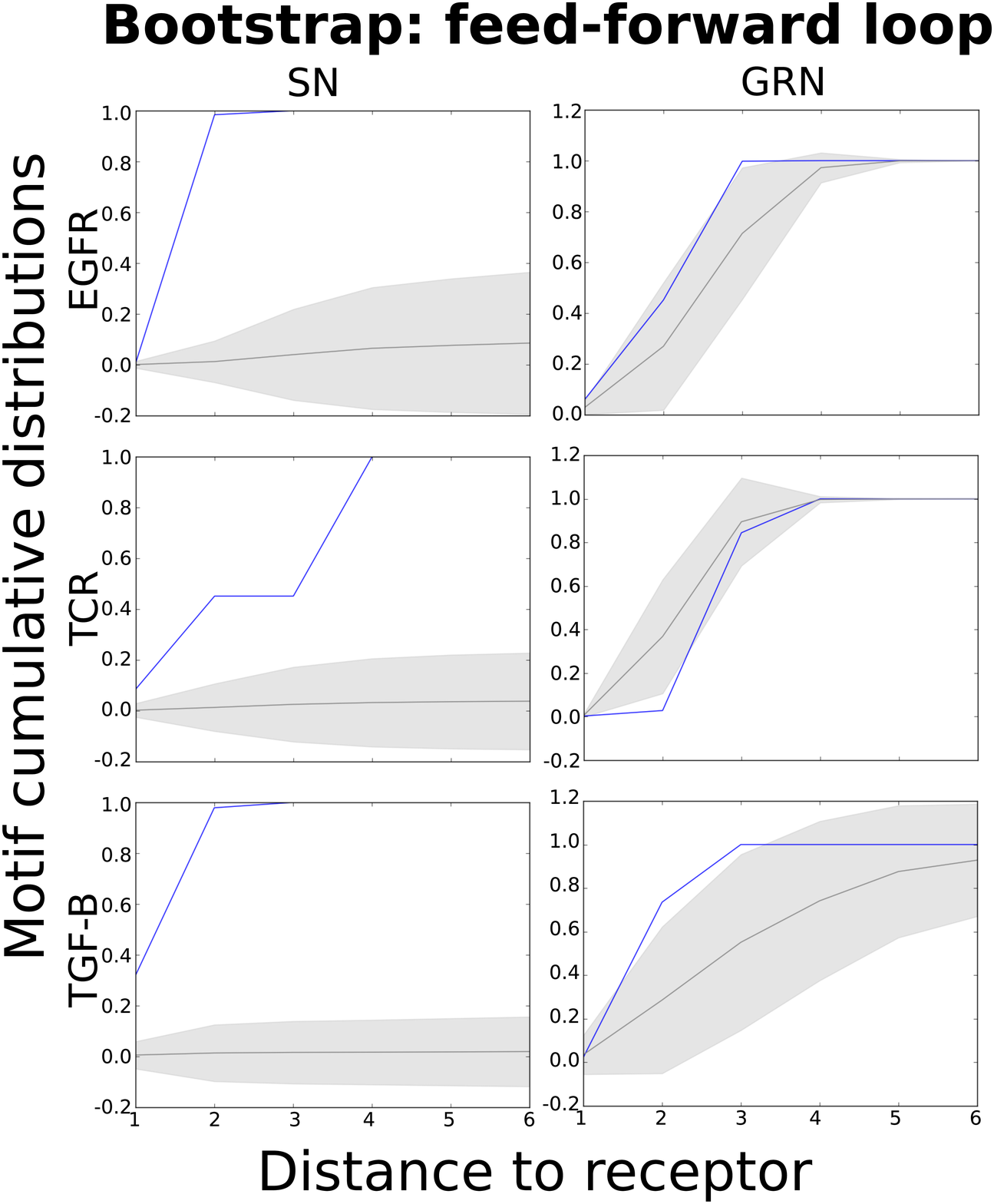}
    \caption{Bootstrap analysis for the feed-forward loop. The blue line shows the observed distribution. The grey line stands for the random ensemble mean, and the shaded area represents a standard deviation above and below the mean.}
    \label{fig:boots}
\end{figure}

We answer this question by comparing the networks' motif cumulative distribution to random networks with the same degree distribution, generated as described in section \ref{sec:netana}. Our results for the feed-forward loop are displayed in Figure \ref{fig:boots} and Table \ref{tab:stats}. It is evident that signalling networks are further removed from the distribution generated by random networks than their gene regulatory counterparts. Following the idea that deviations from the random ensemble indicate natural selection, our results suggest that motif distributions have indeed been tailored to specific pathways -- insofar as signalling components are involved. As for the GRNs, there is not enough evidence to indicate evolutionary pressure; a possible explanation is that ``heavy-duty'' signal processing is performed in signalling pathways (see, for instance, Figure \ref{fig:DIP} and the associated discussion in this section), and GRNs are simply responsible for carrying out the consequences of cellular signal integration. 

Additionally, the feedback loop is given a special place in all networks, being placed further from the upstream node than other motifs. This tendency, already observed in signalling networks by Ma'ayan et al. \cite{Maayan}, is reinforced here and extended to gene regulatory circuitry. As with other principles of cell network evolution, this has plausibly come to be due to distinct dynamical properties for the feedback loop when compared to other motifs. We cite here the roles of the feed-forward and feedback loops in amplifying signals and filtering out noise, respectively, as an example \cite{SysBio,Shin}.

\begin{table}[!h]
\centering
\caption{Bootstrap statistics for the motif distributions. The left number indicates the network's z-score, and the right number its p-value (see Section \ref{sec:netana}).}
\label{tab:stats}
\begin{tabular}{|l|c|c|}
\hline
     & \textbf{SN} & \textbf{GRN} \\
     \hline
    \textbf{TCR} & 9.27; 0.006 & 0.57; 0.22\\
    \hline
    \textbf{EGFR} & 3.42; 0.002 & 0.034; 0.35 \\
    \hline
    \textbf{TGF-\textbeta} & 7.14; 0.008 & -0.14; 0.43\\
\hline
\end{tabular}
\end{table}

\begin{figure}[!h]
    \centering
    \includegraphics[width=.95\columnwidth]{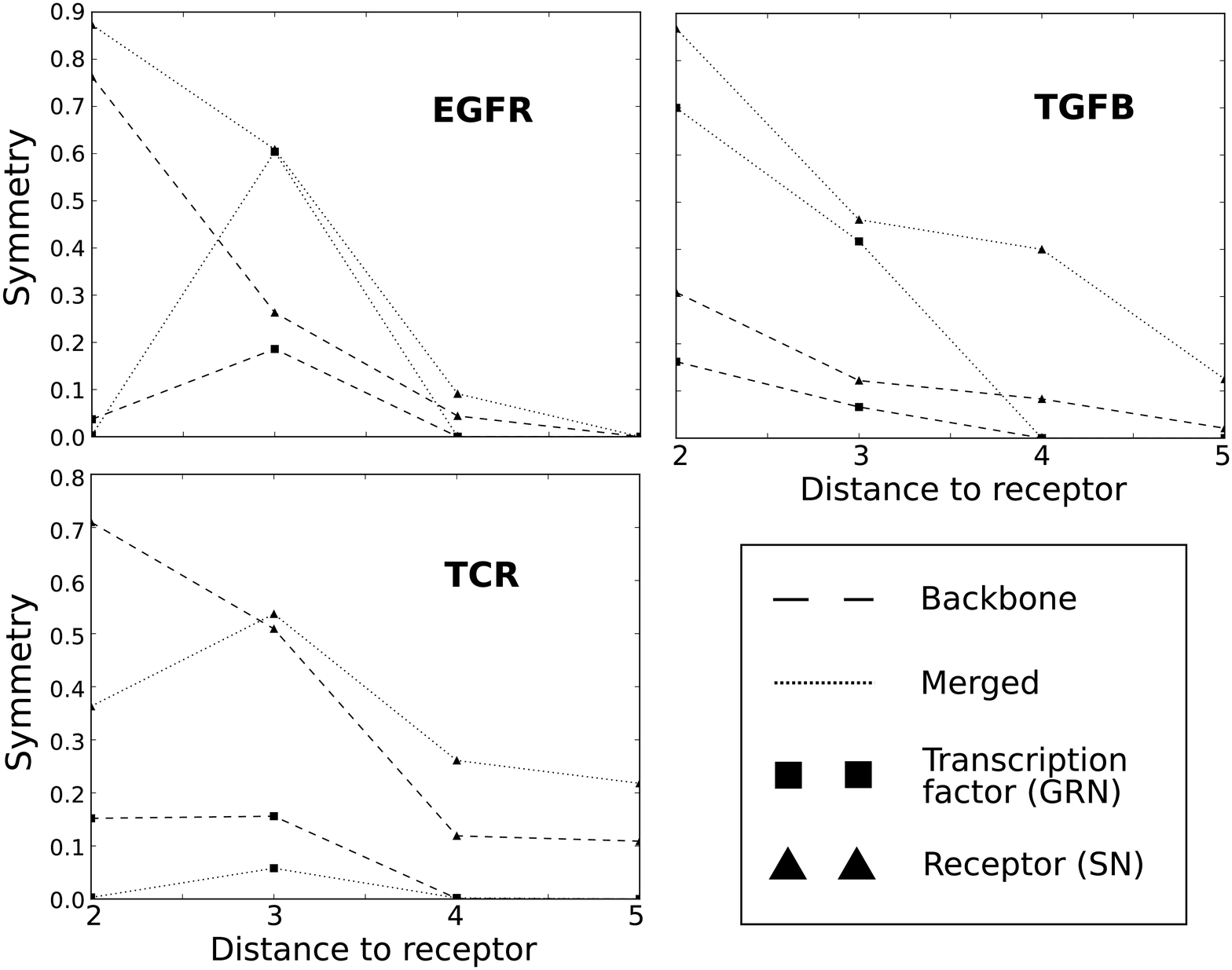}
    \caption{Concentric symmetries for increasing neighbourhoods of the receptor nodes in each studied network. In the lower right corner, a blow-up of the legend for all symmetry plots. The transcription factors and receptors for each pathway are described in the text (see section III).}
    \label{fig:sym}
\end{figure}

With regards to symmetry, it is also observed that signalling networks are considerably more symmetric around their input nodes (Figure \ref{fig:sym}). We see that the cytosolic components consistently display higher symmetry values in neighbourhoods closer to the receptor when compared to corresponding GRN components. Again, this might reflect a different usage of the motifs' dynamical properties by the SNs when compared to GRNs. In particular, the higher symmetries around the receptor node of signalling networks suggest an uniforming constraint on the paths originating at the upstream node, and consequently on the placement of motifs around it.

Next, we compare our results to those of Ma'ayan et al. As discussed in section \ref{sec:netana}, the mCDF and MLI are both related to how motifs are arranged relative to a receptor. However, the MLI considers motif placement relative to an origin and a destination (a cellular machinery) in the network, while the mCDF considers only a starting point. In analysing the MLI distributions for different motifs and machineries (see Figure 5 in reference \cite{Maayan}), we may expand our conclusions by comparing with a related measure. Ma'ayan's analyses place most motifs roughly at a halfway point between receptor and target machinery, regardless of motif or signalling pathway. When compared to our results on the motif cumulative distribution (Figure \ref{fig:MoCDF}), which suggest a vast heterogeneity of motif placements depending on the network and motif in question, we are drawn to the conclusion that different machineries are found at differing distances from the input node. Thus, differences in path length in a signalling network may be related to cellular machinery activation and decision making.

\begin{figure}[!h]
    \centering
    \includegraphics[width=0.95\columnwidth]{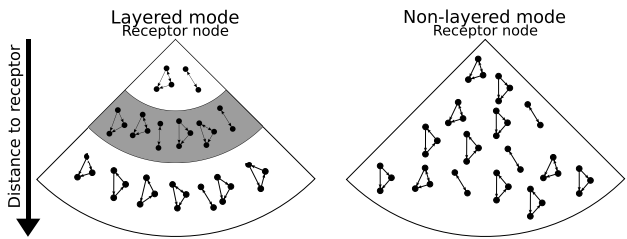}
    \caption{The two types of motif placement found in this work. To the left, the ``motif belt'' observed in SNs; to the right, the asymmetric distribution present in GRNs.}
    \label{fig:did}
\end{figure}

Finally, by combining both aspects of symmetry and cumulative distribution of motifs, as well as previous results in the area \cite{Maayan}, we see the emergence of two distinct patterns of motif placement (see Figure \ref{fig:did}): in signalling pathways, the location of motifs is strictly constrained, leading to concentric ``motif belts'', which may be one or more, around the receptor. For instance, the TCR signalling network shows two belts at distances one and three from its receptor, while the EGFR pathway shows only a single belt three steps away (see Figure \ref{fig:DIP}). Gene regulatory networks, on the other hand, present a more relaxed distribution when compared to random networks with the same degree distribution and suggest a different use for the motifs' dynamical properties.

\begin{figure}[!h]
    \centering
    \includegraphics[width=.9\columnwidth]{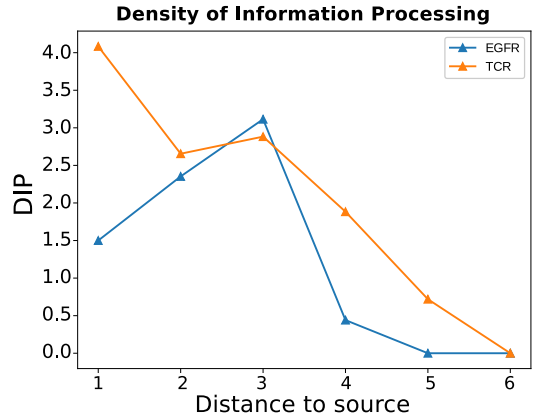}
    \caption{Density of Information Processing for the TCR and EGFR signalling pathways.}
    \label{fig:DIP}
\end{figure}

These findings reassert the notion that cellular networks evolved as modular, relatively independent solutions to distinct selective pressures \cite{Jordan}. Each part of a cell's machinery would then exhibit different organisational features to address their specific demands \cite{Top,AlonEvo}, and as a particular example we find here the difference between motif organisation of signalling networks and gene regulatory networks. As sources of distinct selective pressures, we could cite the different timescales of network dynamics (proteins transition between their active and inactive states much faster than the transcription of genes; see Alon \cite{SysBio}) and the need for SNs to cope with ever-changing external environments \cite{Top,Newman,Kaern}.

\section{Conclusion}
The idea of motifs as regular components and processing units of biological networks is central in our understanding of cellular systems. They have been shown to be ubiquitous in situations as distinct as gene regulation, signal processing and metabolism. Additionally, advances in characterising the relationship between network topology and dynamics point toward special roles of particular motifs such as the feed-forward and feedback loops. Despite these advances in characterising both GRNs and SNs, studies combining both of them remain (to our knowledge) scarce. Here, we compared them with respect to their characteristics pertaining signal transduction and propagation.

Our results add another layer of versatility to the functional importance of motifs by suggesting that their topological distribution differs from signalling to gene regulatory networks. More specifically, signalling pathways show one or more symmetric layers of motifs around the receptor, differing strongly from random networks with the same degree distribution. In contrast, gene regulatory networks display asymmetric motifs in a single layer around key transcription factors, on par with random networks of the same degree distribution. As a remark, feedback loops are usually lagged behind other motifs, as was already noted by Ma'ayan et al.

Thus, our work expands on the previous notion that biological networks in different locations of the cell, or performing different functions, exhibit distinct topological features. Such diversity of topologies could, conceivably, have emerged as separate evolutionary answers to the different selective pressures acting on the cellular components. As such, efforts to understand exactly what the demands of each cellular subnetwork are, and how the cell addresses them, can offer great insights on both the organisation of cellular circuits and the dynamics of network systems.

\section*{Acknowledgements}
The authors acknowledge financial support from FAPESP grants 2015/22308-2, 2012/19278-6 and 2014/19323-7.

\end{document}